\documentclass[9pt,aps,pra,twocolumn,showpacs,reprint,superscriptaddress]{revtex4}
\usepackage{mathrsfs}
\usepackage{amsmath,amssymb,amsfonts}
\usepackage{graphicx}
\usepackage{bm}
\bmdefine\bmzeta{\zeta}
\bmdefine\bmsigma{\sigma}

\begin{document}

\title{Symmetry protected skyrmions in 3D spin-orbit coupled Bose gases}

\author{Guanjun Chen}
\affiliation{Institute of Theoretical Physics, Shanxi University, Taiyuan 030006, China}
\affiliation{Department of Physics, Taiyuan Normal University, Taiyuan 030001, China}
\author{Tiantian Li}
\affiliation{Institute of Theoretical Physics, Shanxi University, Taiyuan 030006, China}
\author{Yunbo Zhang}
\email{ybzhang@sxu.edu.cn}
\affiliation{Institute of Theoretical Physics, Shanxi University, Taiyuan 030006, China}
\date{\today}

\begin{abstract}
We present a variational study of pseudo-spin $1/2$ Bose gases in a
harmonic trap with weak 3D spin-orbit coupling of
$\bmsigma\cdot\mathbf{p}$ type. This spin-orbit coupling
mixes states with different parities, which inspires us to approximate the single particle state
with the eigenstates of the total angular
momentum, i.e. superposition of harmonic $s$-wave and $p$-wave
states. As the time reversal symmetry is protected by
two-body interaction, we set the variational order parameter as
the combination of two mutually time reversal symmetric
eigenstates of the total angular momentum. The variational results
essentially reproduce the 3D skyrmion-like ground state recently
identified by Kawakami {\it et al.}. We show that these
skyrmion-like ground states emerging in this model are primarily
caused by $p$ wave spatial mode involving in the variational order
parameter that drives two spin components spatially separated. We
find the ground state of this system falls into two phases with
different density distribution symmetries depending on the relative
magnitude of intraspecies and interspecies interaction: Phase I has
parity symmetric and axisymmetric density distributions, while Phase
II is featured with special joint symmetries of discrete rotational
and time reversal symmetry. With the increasing interaction strength
the transition occurs between two phases with distinct density
distributions, while the topological 3D skyrmion-like spin texture
is symmetry protected.
\end{abstract}


\maketitle

\section{Introduction}
The experimental realization \cite{lin2011spin,galitski2013spin} of
one-dimensional (1D) spin-orbit(SO) coupling in pseudo-spin $1/2$
Bose gases has stimulated many theoretical works on SO coupling in
cold atom physics. These works range from Raman induced 1D SO
coupling
\cite{PhysRevLett.107.150403,PhysRevLett.108.225301,zheng2013properties,PhysRevA.86.041604,zhai2014degenerate}
that has been realized in cold atoms to more symmetric
two-dimensional (2D) Rashba configuration
\cite{PhysRevLett.105.160403,PhysRevLett.108.010402,PhysRevA.85.023606,PhysRevLett.108.035302,PhysRevA.87.051606,zhai2012spin,PhysRevA.87.031604,PhysRevLett.110.140407,zhai2014degenerate}
that has been extensively studied in condensed matter. In the
absence of harmonic trap, single particle ground states of both
Raman induced and Rashba SO coupling are degenerate, and two-body
interaction selects the generic ground state from the degenerate
manifold determined by the interaction parameters. For example, Wang
{\it et al.} \cite{PhysRevLett.105.160403} found two distinct ground
state phases, namely the plane wave and standing wave(or stripe)
phases, appeared when intraspecies two-body interaction is larger or
smaller than interspecies interaction respectively in homogeneous 2D
Rashba SO coupled pseudo-spin $1/2$ Bose gases. In the presence of a
2D harmonic trap, a more complex phase diagram of Rashba SO coupled
Bose gases with two classes of phases and several subphases in each
was figured out by Hu {\it et al.}
\cite{PhysRevLett.108.010402,PhysRevA.85.023606}.

Now experimental schemes for the realization of Rashba SO coupling
have been proposed such as in Ref. \cite{PhysRevA.84.025602}. On the
other hand, the most symmetric three-dimensional (3D) SO coupling or
Weyl coupling, which even doesn't exist in solid matter, is expected
to be realizable in cold atoms gases and experimental schemes for
that have also been proposed theoretically
\cite{PhysRevLett.108.235301,PhysRevLett.111.125301}. Recently,
Kawakami {\it et al.} \cite{PhysRevLett.109.015301} identified a 3D
skyrmion ground state in 3D SO coupled two-component bosons by
numerically minimizing the Gross-Pitaevskii (GP) energy functional
of the system. They explained the stability of 3D skyrmion ground
state as a result of helical modulation of the order parameter in
the presence of SO coupling. The interaction in their work is
supposed to be SU(2) symmetric. Even though two skyrmion-like ground
states are found to be stabilized in different interaction regimes,
a ground state phase diagram is still absent now. In another work by
Li {\it et al.} \cite{li20123d}, the 3D skyrmion-like
ground state is found to emerge in weak SO coupling regime, while skyrmion
lattice arises in strong SO coupling regime.

In this work, we consider a pseudo-spin 1/2 boson system subject to
3D SO coupling of $\bmsigma\cdot\mathbf{p}$ type in a harmonic trap,
and aim to elucidate the role of interaction in determining the
ground state density and spin texture therein. In Sec. II we
introduce the energy functional for the model in rescaled units
of length, energy, interaction and SO coupling strength. In weak SO
coupling case, the single particle energy levels are essentially
harmonic oscillator-like \cite{anderson2013three}, and SO coupling
will mix states with different parities while keeping the total
angular momentum a conservative. In Sec III we first try to couple
two lowest $s$ and $p$ wave states with the same total angular
momentum $1/2$ into two spinor wave functions with total angular
momentum magnetic quantum number $\pm 1/2$ that are time reversal
state of each other. Then we set the variational order parameter as
superposition of these two states just as has been done in 1D and 2D
cases \cite{PhysRevLett.107.150403,PhysRevLett.105.160403}. Finally,
we calculate the energy functional using the proposed variational
order parameter. In Sec IV the ground state phase diagram is
determined by numerically minimizing the energy functional with
respect to the variational parameters, we illustrate the density and
spin texture for the two phases. Sec. V summarizes our main results.

\section{\label{model}Model}

We consider a pseudo-spin 1/2 boson system confined in a harmonic
trap with a weak Weyl type 3D spin-orbit (SO) coupling
$\bmsigma\cdot\mathbf{p}$. The system is described by its
Gross-Pitaevskii (GP) energy functional under the mean-field
approximation
\begin{equation}
\mathcal {E}=\mathcal {E}_{0}+\mathcal {E}_{int},
\end{equation}
where the single particle part is%
\begin{equation}
\mathcal {E}_{0}=\int d^{3}\mathbf{r}\Psi^{\dag}\left(
\mathbf{r}\right)
\left( \frac{\mathbf{p}^{2}}{2m}+\frac{1}{2}m\omega^{2}r^{2}%
+\lambda\bmsigma\cdot\mathbf{p}\right)  \Psi\left( \mathbf{r}\right)
\label{h0}%
\end{equation}
with $m$ the mass of atoms and $\omega$ the trap frequency.
$\Psi=(\psi_{\uparrow},\psi_{\downarrow})^{T}$ denotes spinor order
parameters for bosons with pseudo spin states $\uparrow,\downarrow$,
$\bmsigma=(\sigma_x,\sigma_y,\sigma_z)$ are the Pauli matrices and
$\lambda$ parameterizes the SO coupling strength. The interaction $\mathcal {E}_{int}$ takes
the usual contact form of $s$-wave scattering interaction
\cite{PhysRevLett.77.3276}. We assume now
\cite{PhysRevLett.105.160403,li20123d,PhysRevLett.108.010402,PhysRevA.85.023606}
the two intraspecies interaction parameters being the same
$g_{\uparrow\uparrow }=g_{\downarrow\downarrow}=g$ and define the
relative magnitude of the interspecies and intraspecies parameters
as $c=g_{\uparrow\downarrow}/g_{\uparrow\uparrow }$. The interaction
part is then
\begin{equation}
\mathcal {E}_{int} =\frac{1}{4}\int d^{3}\mathbf{r}\left(  \left(
g+cg\right)  n^{2}+4\left( g-cg\right)  S_{z}^{2}\right).
\label{hint}%
\end{equation}
In Eq. (\ref{hint}), $n\left(  \mathbf{r}\right)
=n_{\uparrow}\left( \mathbf{r}\right)  +n_{\downarrow}\left(
\mathbf{r}\right)  $ is the particle density and $S_{z}$ is the $z$
component of the spin density $\mathbf{S}=\frac{1}{2} \Psi^{\dagger}
\mathbf{\bmsigma}\Psi$ with $n_{\uparrow,\downarrow}\left(  \mathbf{r}%
\right)  =\left\vert \psi_{\uparrow,\downarrow}\left(
\mathbf{r}\right) \right\vert ^{2}$ the particle densities of
two components, respectively. The corresponding Hamiltonian is time-reversal (TR) symmetric with
time reversal operator defined as $T=-i\sigma_{y}K$ and $K$ denotes
the complex conjugate. The system has length scale of the trapping potential
$l_{T}=\sqrt {\hbar/m\omega}$, energy scale $\hbar\omega$, interaction strength
scale $\hbar\omega l_{T}^{3}/N$, and SO coupling strength
scale $\sqrt{\hbar\omega/m}$. If we further normalize the order parameter to
unity, i.e., $\Psi \rightarrow \sqrt{N/l_{T}^{3}}\Psi$
with $N$ the total particle number in the condensate, the
energy functional per particle is obtained as
\begin{align}
\epsilon &  =\int d^{3}\mathbf{r}\Psi^{\dag}\left(  \mathbf{r}%
\right)  \left\{
-\frac{\nabla^{2}}{2}+\frac{r^{2}}{2}+\lambda\bmsigma\cdot\mathbf{p}
\right\}  \Psi\left(  \mathbf{r}\right) \nonumber\\
&  +\frac{1}{4}\int d^{3}\mathbf{r}\left(  \left(  g+cg\right)
n^{2}+4\left(
g-cg\right)  S_{z}^{2}\right). \label{Efunctional}%
\end{align}

\section{Variational Approach}

In the case of weak SO coupling the single particle energy spectrum
in our system should be harmonic oscillator-like as proposed in Ref.
\cite{li20123d,anderson2013three}. The three dimensional harmonic
oscillator thus proves to be a good choice of the trial wave function, upon
which we may develop our variational method. As can be seen later the
spin-orbit coupling induces transition between eigen states with the same
total angular momentum but different parity, which are mixed into the
variational wave function. The interaction Hamiltonian further couples
the two time-reversal states with different weight factor due to the anisotropic
interaction parameter ratio $c$.

\subsection{Variational order parameter}

The eigenequation of three dimensional harmonic oscillator, $\left( -\frac{\nabla^{2}}{2}
+\frac{r^{2}}{2}\right)\phi=\varepsilon\phi$, has well-known
solutions, with energy eigenvalues
$\varepsilon_{n_rl}=2n_{r}+l+\frac{3}{2}$ and
eigenfunctions $\phi_{n_{r}lm_l}\left(  r,\theta,\varphi\right)  =R_{n_{r}%
l}\left(  r\right)  Y_{lm_l}\left(  \theta,\varphi\right)  $. Here
$n_{r} $ is the radial quantum number, $l$ is the orbital angular
momentum quantum number with $m_l$ its magnetic quantum number,
$R_{n_{r}l}$ is the radial wave function, and $Y_{lm_l}$ is the
spherical harmonics. The Casimir operator $\mathbf{l}^2$ and $\mathbf{s}^2$
for the orbital and spin angular momenta and their $z$-components are all
conservatives in the harmonic oscillator problem. In order to take into account
the spin-orbital coupling term $\bmsigma\cdot\mathbf{p}$, it is
convenient to choose the coupled representation of angular momentum,
i.e. the complete set of commutative operators $\mathbf{l}^2,\mathbf{s}^2,\mathbf{j}^2,j_{z}$
where $\mathbf{j}=\mathbf{l}+\mathbf{s}$ and $j_{z}$ denote the total angular
momentum and its $z$-component, respectively. The eigenfunction
$\phi_{n_{r}lm_l}\left( r,\theta,\varphi\right) $ should be combined with
the spin wave function $\chi_{m_s}$ in the coupled representation as%
\begin{equation}
\phi_{n_{r}ljm_{j}}\left(  r,\theta,\varphi\right) =R_{n_{r}l}\left(
r\right) Y_{jm_{j}}^{l}\left( \Omega\right),
\end{equation}
where $Y_{jm_{j}}^{l}\left( \Omega\right)  =\sum_{m_{l},m_{s}}
C_{lm_l\tfrac{1}{2}m_s}^{jm_j} Y_{lm_l} \chi_{m_s}$ is the
spinor spherical harmonics \cite{varshalovich1988quantum} with
$j=l\pm1/2$ and $C_{lm_l\tfrac{1}{2}m_s}^{jm_j} $ the Clebsch-Gordan coefficients.
In the coupled representation, the ground state wave function has $n_r=l=0$.
This gives a total angular momentum $j=\frac{1}{2}$ with
$m_{j}=\pm\frac{1}{2}$ and the two degenerate ground states are
\begin{equation}
\phi_{00\frac{1}{2}\pm\frac{1}{2}}\left(
\mathbf{r} \right) =R_{00}\left(  r\right)  Y_{\frac{1}{2}\pm\frac{1}{2}%
}^{0}\left(  \Omega\right),
\end{equation}
respectively. Because the SO coupling term breaks the parity
symmetry, it can couple $s$ and $p$ wave states
with the same total angular momentum $\mathbf{j}$ and $j_{z}$
\cite{li20123d}. Keeping these consideration in mind, in the simplest
approximation, we suppose the ground state contains only the lowest $s$
and $p$ wave states with total angular momentum quantum number
$j=\frac{1}{2}$ in presence of the SO coupling term. The state with
$m_{j}=\frac{1}{2}$ takes the form
\begin{equation}
\Phi_{j=\frac{1}{2},m_{j}=\frac{1}{2}}  =  N_{\alpha}\left(  \phi_{00\frac{1}{2}\frac{1}{2}%
}+i\alpha\phi_{01\frac{1}{2}\frac{1}{2}}\right)  \label{superposi}
\end{equation}
where $N_{\alpha}=(1+\alpha^{2})^{-1/2}$, $\alpha$
stands for the relative weight of the $s$ and $p$ orbital modes,
and $i$ in front of $\alpha$ originates from the pure imaginary matrix
element of the SO coupling between the two states in Eq. (\ref{superposi}). This
hypothesis is similar to that appears in Refs. \cite{li20123d} and
\cite{anderson2013three}, and has been verified numerically \cite{li20123d}.
Explicitly this state is a spinor
\begin{equation}
\Phi_{j=\frac{1}{2},m_{j}=\frac{1}{2}}   =N_{\alpha}\left(
\begin{array}
[c]{c}%
R_{00}Y_{00}-i\alpha\sqrt{\frac{1}{3}}R_{01}Y_{10}\\
i\alpha\sqrt{\frac{2}{3}}R_{01}Y_{11}%
\end{array}
\right).\label{basicp}%
\end{equation}
The state with
$m_{j}=-\frac{1}{2}$ takes the form
\begin{equation}
\Phi_{j=\frac{1}{2},m_{j}=-\frac{1}{2}}=N_{\alpha}\left( \phi_{00\frac{1}{2}-\frac{1}{2}%
}+i\alpha\phi_{01\frac{1}{2}-\frac{1}{2}}\right)
\end{equation}
and similarly we have
\begin{equation}
\Phi_{j=\frac{1}{2},m_{j}=-\frac{1}{2}}=N_{\alpha}\left(
\begin{array}
[c]{c}%
-i\alpha\sqrt{\frac{2}{3}}R_{01}Y_{1-1}\\
R_{00}Y_{00}+i\alpha\sqrt{\frac{1}{3}}R_{01}Y_{10}%
\end{array}
\right)\label{basicm}%
\end{equation}
which is nothing but the time reversal of $\Phi_{j=\frac{1}%
{2},m_{j}=\frac{1}{2}}$. In the single particle level,
$\Phi_{j=\frac{1}{2},m_{j}=\pm\frac{1}{2}}$ and any normalized
superposition of them has the same energy thus are ``degenerate''
single particle states, which is similar to degeneracy indicated by
Kramers' theorem in spin-$1/2$ system.

The single particle states exhibit infinite-fold degeneracy and we
expect this degeneracy can be partially resolved by the interaction
which would pick up the ground state from these degenerate states
as in the case of Rashba spin-orbital coupling considered
by Wang and Zhai \cite{PhysRevLett.105.160403}. Since the
interaction doesn't break the time reversal symmetry, the
residual two-fold Kramers degeneracy need to be considered
in the wave function \cite{galitski2013spin}. We
therefore set the variational order parameter as%
\begin{align}
\Psi &  =c_{+}\Phi+c_{-}T\Phi\nonumber\\
&  =\left(
\begin{array}
[c]{c}%
c_{+}\Phi_{\uparrow}-c_{-}\Phi_{\downarrow}^{\ast}\\
c_{+}\Phi_{\downarrow}+c_{-}\Phi_{\uparrow}^{\ast}%
\end{array}
\right)  , \label{variationf}%
\end{align}
with the constraint $c_{+}^{2}+c_{-}^{2}=1$. Here
$\Phi\equiv\Phi_{j=\frac{1}{2},m_{j}=\frac{1}{2}}$ and
$\Phi_{\uparrow,\downarrow}$ are its up and down components.
So far, we have introduced three variational parameters $\alpha, c_+, c_-$ and the
energy functional of Eq. (\ref{Efunctional}) can be calculated
analytically using the proposed order parameter (\ref{variationf}).

\subsection{Energy functional}

We calculate the energy functional on the variational wave function (\ref{variationf}).
The contribution comes from two parts, the single particle and the interaction Hamiltonian.
We notice that for the kinetic and trapping potential terms the nonzero integral contribution
comes from those states with the same parities, while spin orbital
coupling $\bmsigma \cdot \mathbf{p}$ term will mix states with
opposite parities, i.e.
\begin{align}
& \int d^{3}\mathbf{r}\Psi^{\dag}\left(  \mathbf{r}%
\right)  \left\{
-\frac{\nabla^{2}}{2}+\frac{r^{2}}{2}+\lambda\bmsigma\cdot\mathbf{p}
\right\}  \Psi\left(  \mathbf{r}\right) \nonumber\\
&  =N_{\alpha}^2\left[  \left\langle \phi_{00\frac{1}{2}\frac
{1}{2}}\left\vert \left(
-\frac{\nabla^{2}}{2}+\frac{r^{2}}{2}\right) \right\vert
\phi_{00\frac{1}{2}\frac{1}{2}}\right\rangle \right.
\nonumber\\
&  +\alpha^{2}\left\langle
\phi_{01\frac{1}{2}\frac{1}{2}}\left\vert \left(
-\frac{\nabla^{2}}{2}+\frac{r^{2}}{2}\right)  \right\vert
\phi_{01\frac{1}{2}\frac{1}{2}}\right\rangle \nonumber\\
&  \left.  +i2\alpha\left\langle
\phi_{00\frac{1}{2}\frac{1}{2}}\left\vert
\lambda\bmsigma\cdot\mathbf{p}\right\vert \phi_{01\frac{1}{2}\frac{1}{2}%
}\right\rangle \right].\label{h0matrix}%
\end{align}
Here we have used
\begin{align}
&  \left\langle \phi_{00\frac{1}{2}\frac{1}{2}}\left\vert \lambda
\bmsigma\cdot\mathbf{p}\right\vert \phi_{01\frac{1}{2}\frac{1}{2}%
}\right\rangle \nonumber\\
&  =\left\langle \phi_{00\frac{1}{2}-\frac{1}{2}}\left\vert
\lambda
\bmsigma\cdot\mathbf{p}\right\vert \phi_{01\frac{1}{2}-\frac{1}{2}%
}\right\rangle ,\label{sopm}%
\end{align}
which is on account of $\left[  j_{z},\bmsigma\cdot\mathbf{p}\right]
=0$.

It is crucial to calculate the contribution of the spin-orbital coupling
term by means of the irreducible tensor method
\cite{varshalovich1988quantum}. To this end we first introduce the
irreducible form of spin-orbital coupling term. The irreducible
tensor form of momentum operator is \cite{varshalovich1988quantum}
\begin{equation}
p^{\left(  1\right)  }=i\sqrt{2}\frac{1}{r}\left\{  C^{\left(
1\right) }l^{\left(  1\right)  }\right\}  ^{\left(  1\right)
}-i\frac{\partial
}{\partial r}C^{\left(  1\right)  },\label{ptensor}%
\end{equation}
where $C^{\left(  1\right)  }$ and $l^{\left(  1\right)  }$ are
rank-1 irreducible tensors of the unit vector $\hat{\mathbf{r}}$ and
the orbital angular momentum $\mathbf{l}$, and $\left\{  A^{\left(
m\right) }B^{\left(  n\right) }\right\}  ^{\left(  k\right)  }$
defines the rank-$k$ tensor product of rank-$m$ irreducible tensor
$A^{\left( m\right) }$ and rank-$n$ irreducible tensor $B^{\left(
n\right) }$. According to \cite{varshalovich1988quantum}, the dot
product of two arbitrary vectors $\mathbf{A} $ and $\mathbf{B}$ is
related to the tensor product through
$\mathbf{A}\cdot\mathbf{B}=-\sqrt{3}\left\{  A^{\left(  1\right)
}B^{\left( 1\right)  }\right\}  ^{\left(  0\right) }$. In our case,
the radial coordinate $r$ can be separated from the spin and
spherical parts accordingly
\begin{align}
\bmsigma\cdot\mathbf{p} &  =-i\frac{\sqrt{6}}{r}\left\{
\sigma^{\left( 1\right)  }\left\{ C^{\left(  1\right)  }l^{\left(
1\right) }\right\}  ^{\left( 1\right)
}\right\}  ^{\left(  0\right)  }\label{tensorso}\nonumber\\
&  +i\sqrt{3}\frac{\partial}{\partial r}\left\{  \sigma^{\left(
1\right) }C^{\left(  1\right)  }\right\}  ^{\left(  0\right)  },
\end{align}
such that
\begin{align}
&  \left\langle \phi_{00\frac{1}{2}\frac{1}{2}}\left\vert
\mathbf{\bmsigma
\cdot p}\right\vert \phi_{01\frac{1}{2}\frac{1}{2}}\right\rangle \notag \\
&  =-i\sqrt{6}\left\langle R_{00}\left(  r\right)  \left\vert \frac{1}%
{r}\right\vert R_{01}\left(  r\right)  \right\rangle \notag \\
&  \times\left\langle Y_{\frac{1}{2}\frac{1}{2}}^{0}\left(
\Omega\right) \left\vert \left\{  \sigma^{\left(  1\right)  }\left\{
C^{\left( 1\right)  }l^{\left(  1\right)  }\right\}  ^{\left(
1\right)
}\right\}  ^{\left(  0\right)  }\right\vert Y_{\frac{1}{2}\frac{1}{2}}%
^{1}\left(  \Omega\right)  \right\rangle \notag \\
&  +i\sqrt{3}\left\langle R_{00}\left(  r\right)  \left\vert \frac{d
}{d r}\right\vert R_{01}\left(  r\right)  \right\rangle \notag  \\
&  \times\left\langle Y_{\frac{1}{2}\frac{1}{2}}^{0}\left(
\Omega\right) \left\vert \left\{  \sigma^{\left(  1\right)
}C^{\left( 1\right) }\right\}  ^{\left(  0\right)  }\right\vert
Y_{\frac{1}{2}\frac {1}{2}}^{1}\left(  \Omega\right) \right\rangle. 
\label{somatrix}%
\end{align}
The integrals for the radial coordinates are easy to calculate,%
\begin{equation}
\left\langle R_{00}\left(  r\right)  \left\vert
\frac{1}{r}\right\vert
R_{01}\left(  r\right)  \right\rangle =\sqrt{\frac{2}{3}},\label{radial1}%
\end{equation}
\begin{align}
&  \left\langle R_{00}\left(  r\right)  \left\vert \frac{d }{d
r}\right\vert R_{01}\left(  r\right)
\right\rangle  =-\frac{1}{\sqrt{6}},\label{radial2}%
\end{align}
where $R_{00}\left(  r\right)  =\sqrt{2^{2}/\sqrt{\pi}}e^{-r^{2}/2}$
and
$R_{01}\left(  r\right)  =\sqrt{2^{3}/\left(  3\sqrt{\pi}\right)  }%
re^{-r^{2}/2}$ are used. Wigner--Eckart theorem can be used to
calculate the angular and spin integral
\begin{equation}
\left\langle Y_{\frac{1}{2}\frac{1}{2}}^{0}\left( \Omega\right)
\left\vert \left\{  \sigma^{\left(  1\right)  }\left\{ C^{\left(
1\right) }l^{\left(  1\right)  }\right\}  ^{\left( 1\right)
}\right\}  ^{\left( 0\right)  }\right\vert
Y_{\frac{1}{2}\frac{1}{2}}^{1}\left(
\Omega\right)  \right\rangle =\frac{1}{\sqrt{6}},\label{angular1}%
\end{equation}
\begin{equation}
\left\langle Y_{\frac{1}{2}\frac{1}{2}}^{0}\left( \Omega\right)
\left\vert \left\{  \sigma^{\left(  1\right) }C^{\left(  1\right)
}\right\} ^{\left(  0\right)  }\right\vert
Y_{\frac{1}{2}\frac{1}{2}}^{1}\left(  \Omega\right)  \right\rangle =\frac{1}{\sqrt{3}}.\label{angular2}%
\end{equation}
Substitute Eq. (\ref{radial1}-\ref{angular2}) into Eq.
(\ref{somatrix}), one has
\begin{equation}
\left\langle \phi_{00\frac{1}{2}\frac{1}{2}}\left\vert
\bmsigma\cdot\mathbf{p}\right\vert
\phi_{01\frac{1}{2}\frac{1}{2}}\right\rangle =-i\sqrt
{\frac{3}{2}}.\label{somatrix2}%
\end{equation}
Hence the single particle part of the energy functional is%
\begin{align}
& \int d^{3}\mathbf{r}\Psi^{\dag}\left(  \mathbf{r}%
\right)  \left\{
-\frac{\nabla^{2}}{2}+\frac{r^{2}}{2}+\lambda\bmsigma\cdot\mathbf{p}
\right\}  \Psi\left(  \mathbf{r}\right) \nonumber\\
&  =N_{\alpha}^2\left( \frac{3}{2}+\frac{5}{2}\alpha^{2}+\sqrt
{6}\alpha\lambda\right)  \label{single}%
\end{align}
where we have used the eigen energies of the $s$ and $p$ states of
the three dimensional oscillator are respectively
$\varepsilon_{00}=3/2$ and $\varepsilon_{01}=5/2$.

For the calculation of the interaction part of energy functional, it is easy
to show that the total density $n$ is always spherical symmetric
\begin{align}
n &  =\left\vert \Phi\right\vert ^{2} =(4\pi)^{-1}N_{\alpha}^2\left(
R_{00}^{2}+\alpha
^{2}R_{01}^{2}\right),  \label{density}%
\end{align}%
and the density-density interaction energy is
\begin{equation}
\int
d^{3}\mathbf{r}n^{2}=\tfrac{1}{12}N_{\alpha}^4(2\pi)^{-\tfrac{3}{2}}(5\alpha^{4}+12\alpha^{2}+12)
.\label{densityinter}%
\end{equation}
On the other hand, the spin density is anisotropic, e.g. the $z$ component takes the form of
\begin{align}
S_{z} &=(8\pi)^{-1}N_{\alpha}^2 \left\{ (c_{+}^{2}-c_{-}^{2})(
R_{00}^{2}+\alpha^{2}R_{01}^{2}\cos2\theta)\right.
\nonumber\\
& \left. -(2c_{+}c_{-})( 2\alpha R_{00}R_{01}\sin\theta\sin\varphi
-\alpha^{2}R_{01}^{2}\sin2\theta\cos\varphi)\right\},
\end{align}%
and the spin-spin interaction energy is integrated as
\begin{align}
 \int
d^{3}\mathbf{r}S_{z}^{2}&=\tfrac{1}{4}N_{\alpha}^4(2\pi)^{-\tfrac{3}{2}} \left[  \tfrac{1}{36}(7\alpha^{4}
-12\alpha^{2}+36) \right.\nonumber\\
& \left. -\tfrac{1}{3}c_{+}^{2}c_{-}^{2}%
(\alpha^{4}-12\alpha^{2}+12)\right].  \label{spininter}%
\end{align}
Collecting Eqs. (\ref{single}), (\ref{densityinter}) and (\ref{spininter})
into Eq. (\ref{Efunctional}), we finally arrive at the variational result for
the ground state energy per particle
\begin{align}
\epsilon &  =N_{\alpha}^2\left(  \frac{3}{2}+\frac{5}{2}%
\alpha^{2}+\sqrt{6}\alpha\lambda\right)  \nonumber\\
&
+N_{\alpha}^4(2\pi)^{-\tfrac{3}{2}}\frac{1}{72}\left[(11\alpha^{4}+12\alpha^{2}+36)
g+(4\alpha^{4}+24\alpha^{2})cg \right.  \nonumber\\
& \qquad \qquad  \qquad \left. -6 c_{+}^{2}c_{-}^{2}
\left(g-cg\right)(\alpha^{4}-12\alpha^{2}+12)\right].  \label{energy}%
\end{align}

\section{Ground state phase diagram}
 \begin{figure}
 \includegraphics[width=0.8\columnwidth]{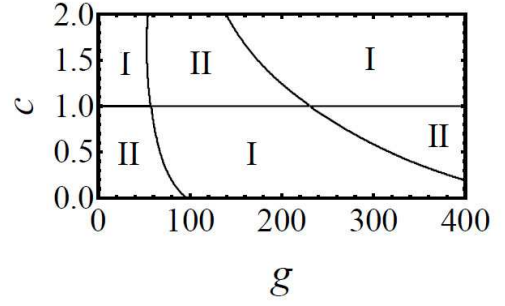}%
 \caption{\label{fig1}Phase diagram of weakly SO coupled two-component Bosons with 
 coupling strength $\lambda=0.2$, which shows two skyrmion-like phases
I and II. Phase I is a skyrmion ground
state of order parameter $\exp\left[ -i\mathbf{\Omega}\left(
\mathbf{r}\right) \cdot \mathbf{S}\right]\bmzeta$ with
$\bmzeta_z=(1,0)^{T}$ and Phase II is a skyrmion state with
$\bmzeta_x=\tfrac{1}{\sqrt{2}}(1,1)^{T}$. Density distribution and
spin texture of these two phases are shown in Fig. \ref{phase1} and Fig.
\ref{phase2} respectively.}
 \end{figure}
The ground state phase diagram can be determined numerically via the
minimization of the variational energy with respect to the
parameters $\alpha$, $c_{+}$ and $c_{-}$ for given $c$ and $g$. We
notice that the parameters $c_{+}$ and $c_{-}$ appear only in the
last term of Eq. (\ref{energy}) in the form of $c_{+}^{2}c_{-}^{2}$,
the value of which ranges from $0$ to $1/4$. The parameter
$c_{+}^{2}c_{-}^{2}$ as a whole takes the value of either $0$ or
$1/4$ in the minimization, depending on the signs of $(g-cg)$ and
$f(\alpha)=\alpha^{4}-12\alpha^{2}+12$. The ground state thus falls
into two classes of phases as depicted in FIG.~\ref{fig1}: Phase I,
the variation yields $\left\vert c_{+}\right\vert ^{2}=1$,
$\left\vert c_{-}\right\vert ^{2}=0$ or $\left\vert c_{+}\right\vert
^{2}=0$, $\left\vert c_{-}\right\vert ^{2}=1$; Phase II, the
variation yields $\left\vert c_{+}\right\vert ^{2}=\left\vert
c_{-}\right\vert ^{2}=1/2$. It is clear that $\alpha$ must be
negative for a positive $\lambda$ due to the fact that $\alpha$'s in
Eq. (\ref{energy}) are all even-ordered except the spin-orbital
coupling term. We see that $c=1$ divides the phase plane into upper
and lower regions. With increasing $g$ the system enters alternately
into Phases I and II and the boundaries are determined by
$f(\alpha)=0$, i.e. $\alpha_{\pm}=- \sqrt{6\pm 2\sqrt{6}}$. For
typical experiments with $^{87}Rb$ condensate, the interaction
strength scale is $10^{-13}$Hzcm$^3$, which gives rise to $g\sim
40-80$. For smaller $g$, $\alpha\ge \alpha_{-}$ results in a
positive value of $f(\alpha)$ such that $c>1$ region belongs to
Phase I and $c<1$ belongs to Phase II. By adjusting the trapping
frequency and the density of the condensate one can easily increase
$g$ to cross the critical line such that $\alpha \in
[\alpha_{+},\alpha_{-}]$, which makes $f(\alpha)$ negative. We
observe an interesting swap of the phases: $c<1$ corresponds to
Phase I and $c>1$ corresponds to Phase II. Similar phase transition
appears in the Rashba spin-orbital coupled Bosons
\cite{PhysRevLett.108.010402,PhysRevA.85.023606}. Further increasing
the interaction strength makes the optimized parameter $\alpha \leq
\alpha_{+}$ and the phases swap occurs again.

 \begin{figure}
 \centering{
 \includegraphics[width=0.8\columnwidth]{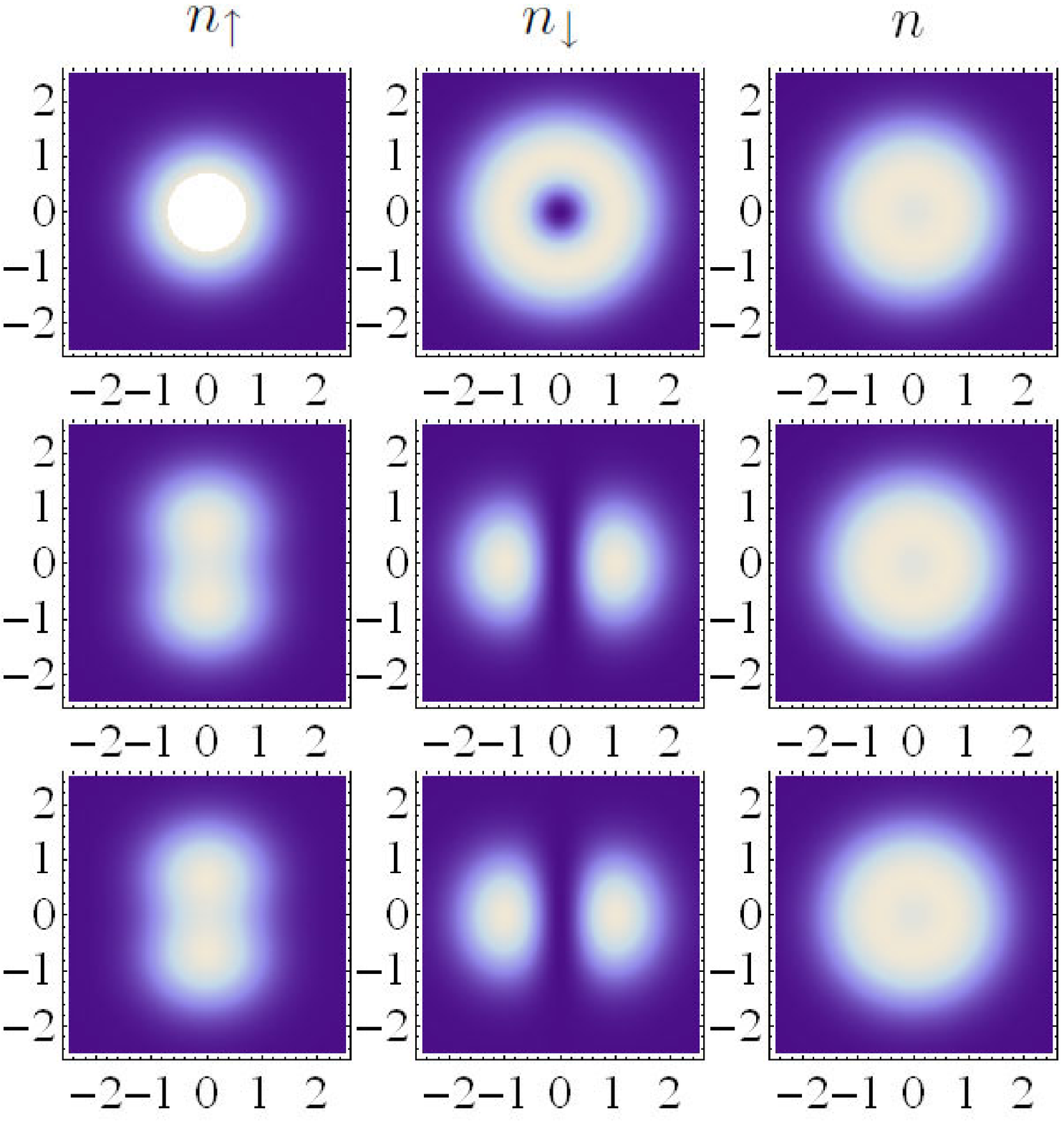}\\
 \includegraphics[width=0.8\columnwidth]{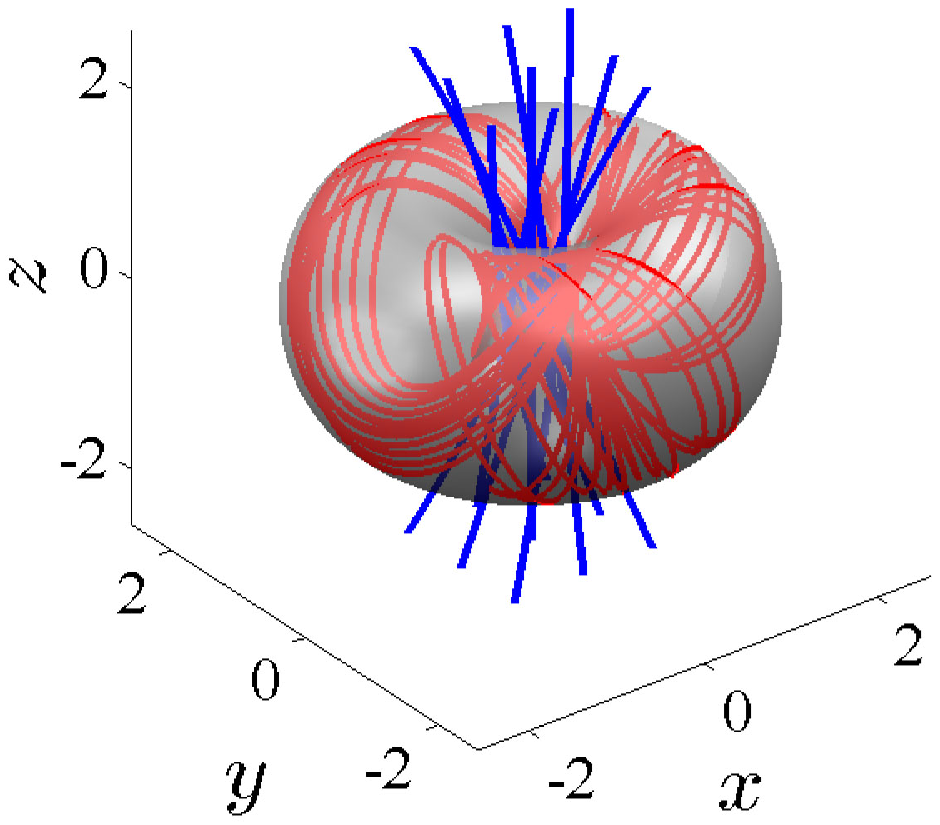}}
 \caption{\label{phase1}(Color online). Density distribution and spin
texture of Phase I for $\alpha=-1.6$. Top: Three rows are densities in $xy$, $yz$ and $xz$ planes
respectively, three columns are for up, down
components and the total density as explicitly labeled above each
column. Density distributions in $xz$ and $yz$
plane are the same due to the $z$-axis rotational symmetry. 
Bottom: 3D skyrmion spin texture
$\mathbf{s}(\mathbf{r})=\mathbf{S}(\mathbf{r})/n(\mathbf{r})$ of
Phase I. The streamline plot of $\mathbf{s}$ in a selected region
are shown.}
 \end{figure}

\begin{figure}
 \centering{
 \includegraphics[width=0.8\columnwidth]{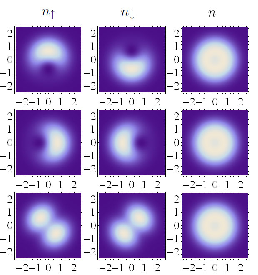}\\
 \includegraphics[width=0.8\columnwidth]{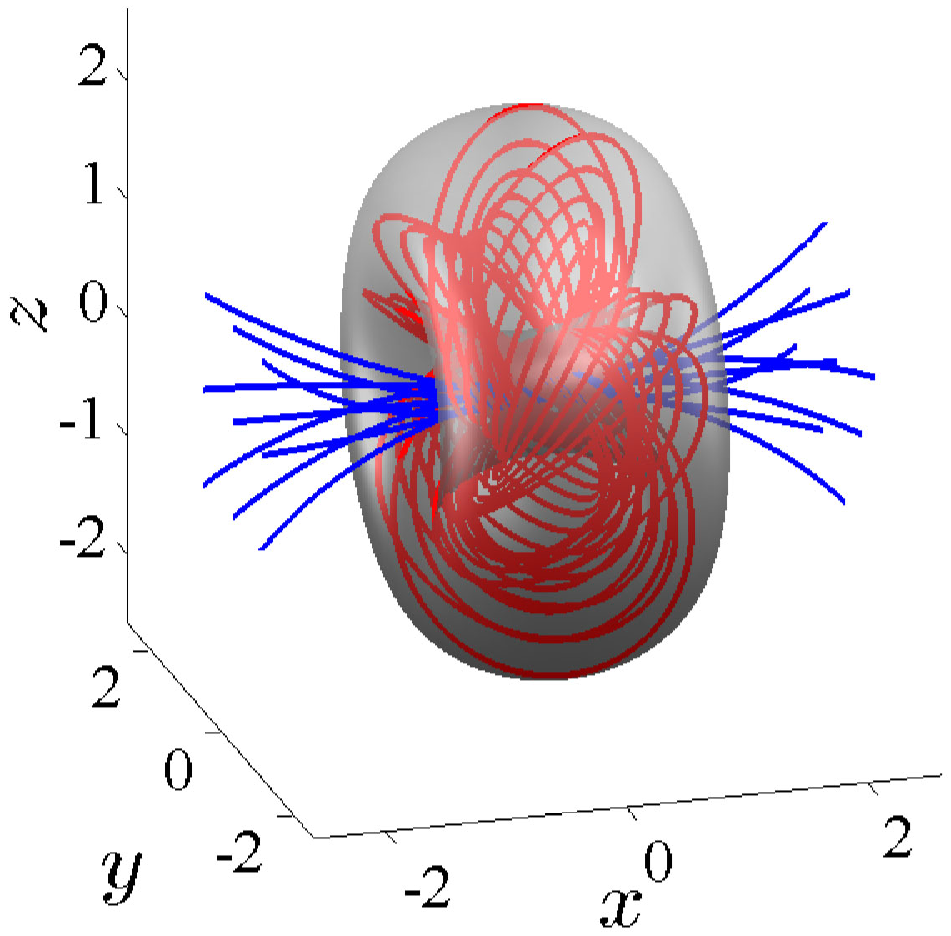}}
 \caption{\label{phase2}(Color online). Density distribution and spin
texture of Phase II for $\alpha=-1.6$. Top: Three rows are densities in 
$xy$, $yz$ and $xz$ planes
respectively, three columns are for up, down
components and the total density as explicitly labeled above each
column. Though the total density is isotropic again, the density distribution
for the two components exhibits more complex symmetry as described in the text.
Bottom: 3D skyrmion spin texture
$\mathbf{s}(\mathbf{r})=\mathbf{S}(\mathbf{r})/n(\mathbf{r})$ of
Phase II, which is roughly a $\pi/2$ rotation about $y$-axis of that 
in Phase I. The topological structure of the spin texture is protected by the 
time reversal symmetry.}
 \end{figure}

The density distributions and the spin texture of Phases I and II
are shown in Figs. \ref{phase1} and \ref{phase2}, respectively.
Typical features include:

{\it Phase I}: this phase contains two degenerate
states $\left\vert c_{+}\right\vert ^{2}=1$, $\left\vert
c_{-}\right\vert ^{2}=0$  and $\left\vert c_{+}\right\vert ^{2}=0$,
$\left\vert c_{-}\right\vert ^{2}=1$. They are time reversal states of each
other and have similar density and
spin texture except that the spin-up and spin-down components are exchanged.
The order parameter for the former has the form
\[
\Psi=(4\pi)^{-\frac{1}{2}}N_{\alpha}\left(
\begin{array}
[c]{c}%
R_{00}\left(  r\right)  -i\alpha R_{01}\left(  r\right)  \cos\theta\\
-i\alpha R_{01}\left(  r\right)  \sin\theta e^{i\varphi}%
\end{array}
\right).
\]
The particle densities for the spin-up and spin-down components are
\begin{align}
n_{\uparrow} &  =(4\pi)^{-1}N_{\alpha}^2\left(  R_{00}%
^{2}\left(  r\right)  +\alpha^{2}R_{01}^{2}\left(  r\right)  \cos^{2}%
\theta\right),  \nonumber\\
n_{\downarrow} &  =(4\pi)^{-1}N_{\alpha}^2\alpha^{2}%
R_{01}^{2}\left(  r\right)  \sin^{2}\theta, \label{nI}
\end{align}
which respect the rotational symmetry about $z$-axis and the parity
symmetry, i.e.
$n_{\uparrow,\downarrow}(\mathbf{r})=n_{\uparrow,\downarrow}(r,\theta)$
and
$n_{\uparrow,\downarrow}(r,\theta)=n_{\uparrow,\downarrow}(r,\pi-\theta)$.
The densities of the two components in $xz$ and $yz$ planes are the same
as shown in Fig. \ref{phase1}, which exhibit clearly characters of the $p$ wave
state, i.e. the spin-up component is dumbbell-like while the spin-down
component forms a torus. The total density on the other hand is isotropic - the sum
of $n_{\uparrow}$ and $n_{\downarrow}$ in Eq. (\ref{nI}) relies only on the radius $r$.

This spin density calculated on the variational order parameter shows interesting
spin texture described by%
\begin{align}
S_{x}& =(4\pi)^{-1}N_{\alpha}^2\left( \alpha R_{00}\left(  r\right)
R_{01}\left(  r\right)  \sin\theta\sin\varphi\right.  \nonumber\\
&  \left.  +\alpha^{2}R_{01}^{2}\left(  r\right)
\sin\theta\cos\theta
\cos\varphi\right),  \nonumber\\
S_{y}& =(4\pi)^{-1}N_{\alpha}^2\left( -\alpha R_{00}\left(  r\right)
R_{01}\left(  r\right)  \sin\theta\cos\varphi\right.  \nonumber\\
&  \left.  +\alpha^{2}R_{01}^{2}\left(  r\right)
\sin\theta\cos\theta
\sin\varphi\right),  \nonumber\\
S_{z}& =(8\pi)^{-1}N_{\alpha}^2\left( R_{00}^{2}\left( r\right)
+\alpha^{2}R_{01}^{2}\left(  r\right) \cos2\theta\right).
\end{align}
The average value of the spin in the $xy$ plane is zero, i.e.
$\langle S_{x}\rangle=\langle S_{y}\rangle=0$. The spin texture
$\mathbf{s}(\mathbf{r})=\mathbf{S}(\mathbf{r})/n(\mathbf{r})$ is
depicted in Fig. \ref{phase1} and we find that spin density forms a
torus near the $xy$ plane and a bundle of nearly vertical
streamlines of spin penetrate the central region of the torus. This
skyrmion-like texture has been discussed in Ref.
\cite{PhysRevLett.109.015301} and identified as the ground state in
$c<1$ regime for an interaction parameter $c_0=100$. Li {\it et al}.
\cite{li20123d} also found this ground state skyrmion spin texture
in weak SO coupling case for isotropic interaction $c=1$. The term
``skyrmion-like'' means the absence of boundary condition at $r
\rightarrow \infty$ \cite{PhysRevLett.109.015301} thus the winding
number for the texture is not an integer.

In order to get a deep understanding of the skyrmion nature of this
ground state, we notice that the order parameter can be obtained
from a local spin rotation from the polarized spinor wavefunction
$\bmzeta_z=(c_{+},c_{-})^{T}=(1,0)^{T}$
\begin{align}
\Psi_{z} &  =\exp\left(  -i\mathbf{\Omega}\left(  \mathbf{r}\right)
\mathbf{\cdot s}\right)  \sqrt{n\left(  \mathbf{r}\right)  }\bmzeta
_{z}
\end{align}
supposing that $n\left(  \mathbf{r}\right)
=(4\pi)^{-1}N_{\alpha}^2\left( R_{00}^{2}\left(  r\right)  +
\alpha^2 R^2_{01}\left(  r\right) \right)  $ and
$\mathbf{\Omega}\left( \mathbf{r}\right) =\omega\left( r\right)
\mathbf{r/}r$. This operation rotates the spin at position
$\mathbf{r}$ by an angle $\omega\left( r\right) $ about the axis
$\mathbf{r/}r$. The rotation angle is position dependent, i.e.
$\omega\left(  r\right) =2\arctan \left( \alpha R_{01}\left(
r\right) /R_{00}\left(  r\right)  \right)$ and $\mathbf{s}$ is the
usual spin angular momentum operators for spin-1/2. It is the
explicit form of $\mathbf{\Omega}\left(\mathbf{r}\right) $ that
determines the specific texture of the skyrmion
\cite{PhysRevA.64.043612}. The polarized spinor order parameter
$\bmzeta_{z}$ has all spins being oriented in the positive
$z$-direction. After the rotation the order parameter $\Psi_z$ for
this skyrmion state is position-dependent. The order parameter is
the most symmetrically shaped skyrmion with the symmetric axis
unrotated, which is identical with that already discussed in Refs.
\cite{al2001skyrmions,PhysRevA.64.043612,PhysRevLett.86.3934,
PhysRevLett.91.010403,PhysRevLett.97.080403,PhysRevA.79.053626}. In
those papers the 3D skyrmion states are proposed as excited states
in pseudo-spin $1/2$ or ferromagnetic spin-1
\cite{PhysRevA.64.043612} Bose gases although they may be
metastable.

{\it Phase II}: this phase again contains two degenerate states with
$c_{+}=c_{-} =\pm\frac{1}{\sqrt{2}}$ and
$c_{+}=-c_{-}=\pm\frac{1}{\sqrt{2}}$, which are time reversal states
of each other. They share similar density distribution and
spin texture just like the case of Phase I. The order parameter for $c_{+}%
=c_{-}$ has the form
\begin{align}
\Psi & =(8\pi)^{-\frac{1}{2}}N_{\alpha} \left(
\begin{array}
[c]{c}%
R_{00}\left(  r\right)  -i\alpha R_{01}\left(  r\right)  (\cos\theta
+\sin\theta e^{-i\varphi})\\
R_{00}\left(  r\right)  -i\alpha R_{01}\left(  r\right) (-\cos\theta
+\sin\theta e^{i\varphi})
\end{array}
\right),
\end{align}
and the particle densities for the two components are%
\begin{align}
n_{\uparrow} &  =(8\pi)^{-1}N_{\alpha}^2\left(  R_{00}%
^{2}\left(  r\right)  +\alpha^{2}R_{01}^{2}\left(  r\right)  \right.
\nonumber\\
&  \left.  -2\alpha R_{00}R_{01}\sin\theta\sin\varphi+\alpha^{2}R_{01}^{2}%
\sin2\theta\cos\varphi\right),  \nonumber\\
n_{\downarrow} &  =(8\pi)^{-1}N_{\alpha}^2\left( R_{00}^{2}\left(
r\right)  +\alpha^{2}R_{01}^{2}\left(  r\right) \right.
\nonumber\\
&  \left.  +2\alpha R_{00}R_{01}\sin\theta\sin\varphi-\alpha^{2}R_{01}^{2}%
\sin2\theta\cos\varphi\right).\label{p2n}
\end{align}
The density for each component consists of two parts, one is
isotropic that is common for both components, the other is
complementary to each other as shown in the second and fourth lines
in Eq.~ (\ref{p2n}). This leads again to an isotropic total density.
The overall density distribution of the two components can be
visualized as two cashew nuts perpendicularly crossing and partially
overlapping with each other. The distributions in $xy$, $yz$ and
$xz$ planes are shown in Fig. \ref{phase2}. The density
distributions have the following symmetries, e.g., the densities of
two components are invariant under the combined operation of time
reversal and $\pi$ rotation about $x$(or $z$) axis, i.e.
$n_{\uparrow}(r,\pi-\theta,2\pi-\phi)=n_{\downarrow}(r,\theta,\phi)$,
$n_{\uparrow}(r,\theta,\pi+\phi)=n_{\downarrow}(r,\theta,\phi)$,
while the $\pi$ rotation about $y$ axis itself leaves the density
distributions unchanged, i.e.
$n_{\uparrow,\downarrow}(r,\pi-\theta,\pi-\phi)=n_{\uparrow,\downarrow}(r,\theta,\phi)$.

The spin texture associated with the order parameter is expressed as
\begin{align}
S_{x}&
=(8\pi)^{-1}N_{\alpha}^2\times\nonumber\\
&  \left(  R_{00}^{2}+\alpha^{2}R_{01}^{2}\left(  \sin^{2}\theta\cos
2\varphi-\cos^{2}\theta\right)  \right),  \nonumber\\
S_{y}&
=(8\pi)^{-1}N_{\alpha}^2\times\nonumber\\
&  \left(  2\alpha R_{00}R_{01}\cos\theta+\alpha^{2}R_{01}^{2}\sin^{2}%
\theta\sin2\varphi\right),  \nonumber\\
S_{z}&
=(8\pi)^{-1}N_{\alpha}^2\times\nonumber\\
&  \left(  -2\alpha R_{00}R_{01}\sin\theta\sin\varphi+\alpha^{2}R_{01}^{2}%
\sin2\theta\cos\varphi\right).
\end{align}
The average spin polarization along $z$ axis is zero, i.e.
$\langle S_{z}\rangle=0$. The spin texture
$\mathbf{S}(\mathbf{r})/n(\mathbf{r})$ is presented in Fig.
\ref{phase2}. The spin density in this case forms a
torus near the $yz$ plane and the fountain-like streamlines of spin
pass through the hole of the torus, which is more or less like a $\pi/2$ rotation
of the torus in Phase I about $y$ axis.
Similarly, this ground state can be obtained from a
local spin rotation from the spinor order
parameter $\bmzeta_x=(c_{+},c_{-})^{T}=\tfrac{1}{\sqrt{2}}(1,1)^{T}$
that describes a system with all spins pointing to the positive $x$
direction, i.e.
\begin{align}
\Psi_{x} &  =\exp\left(  -i\mathbf{\Omega}\left(  \mathbf{r}\right)
\mathbf{\cdot s}\right)  \sqrt{n\left(  \mathbf{r}\right)  }\bmzeta
_{x}.
\end{align}
The spinor wavefunction $\bmzeta_x$ is related to $\bmzeta_z$
by a $\pi/2$ rotation around $y$. Owing to the non-Abelian
nature of SO$(3)$ rotation, the spin texture of Phase II is
different from the $\pi/2$ rotation around $y$ of Phase I. The difference
between these two textures lies in the fact that the spin in the torus of
Phase I revolves the $z$ axis following an elliptical (oval) orbits that
rotate gradually like the perihelion precession in celestial mechanics,
while in Phase II the orbits are closed loops.
Apart from this, they indeed share the same topology determined by the
same $\mathbf{\Omega}$ as can be seen from the spin streamline plot
in Fig. \ref{phase2}, because the topological spin texture is
protected by time reversal symmetry of the system. This skyrmion spin
texture is proposed as the ground state in the regime of $c>1$ in
Ref. \cite{PhysRevLett.109.015301}.

We find in this study that in both phase I and phase II, the densities
of the two components are spatially separated in three
dimensions. We thus come up with a conclusion that phase
separation of the spin components generally exists in 1D\cite{PhysRevA.90.043619},
2D\cite{PhysRevLett.108.010402,PhysRevA.85.023606} and 3D SO coupled
Boson gases. In our case it is the SOC-induced $p$ wave
spatial mode involving in the variational order parameter that
drives the two spin components spatially separated.
As pointed out by Battye {\it et al.} \cite{PhysRevLett.88.080401}
phase separation is a prerequisite for existence of stable skyrmion, which
explains why skyrmion spin texture appears in the variational ground state of our model.
Furthermore the topology of the skyrmion texture is protected by the time reversal
symmetry of the system even the phase transition drastically changes the
density structure.

\section{Summary}
We have investigated variationally the ground state phase diagram of weakly
3D spin-orbital coupled two-component Bose gases in a harmonic trap.
Two phases for the ground state are identified depending on
intraspecies and interspecies interaction strength and the corresponding density
distribution and the spin texture are illustrated for optimized variational
parameters. Phase I is featured with the parity symmetric and rotational symmetric density
distribution of both spin-up and spin-down components and skyrmion spin texture
with torus in the $xy$ plane and spin streamline passing through the central region, while
Phase II is characterized with density distribution possessing discrete $\pi$ rotational
symmetry about $y$ axis and $\pi$ rotational-time-reversal symmetry about $x$ and $z$ axis
and the similar spin torus is in the $yz$ plane, roughly a $\pi/2$ rotation of that in Phase I about $y$ axis.
In both phases, the density of two components is essentially phase separated. With
increasing interaction strength, interesting phase
transition occurs between the two phases, while the topology of ground
state spin textures is protected.

\begin{acknowledgments}
This work is supported by NSF of China under Grant Nos. 11234008 and
11474189, the National Basic Research Program of China (973 Program) under
Grant No. 2011CB921601, Program for Changjiang Scholars and Innovative
Research Team in University (PCSIRT)(No. IRT13076).
\end{acknowledgments}


\end{document}